\documentclass[preprint2]{aastex63}

\usepackage{soul,xcolor}

\newcommand{\noopsort}[1]{}

\received{XXX}
\revised{YYY}
\accepted{ZZZ}
\submitjournal{ApJ}

\shorttitle{Electron acceleration in turbulence}
\shortauthors{Trotta et al.}

\begin{document}

\title{Fast acceleration of transrelativistic electrons in astrophysical turbulence}

\correspondingauthor{Domenico Trotta}
\email{d.trotta@qmul.ac.uk}

\author[0000-0002-0786-7307]{Domenico Trotta}
\affiliation{School of Physics and Astronomy, Queen Mary University of London, Mile End Road, London E1 4NS, United Kingdom}
\affiliation{Dipartimento di Fisica, Università della Calabria, Ponte P. Bucci, cubo 31C, 87036 Rende, Italy}

\author[0000-0002-7419-0527]{Luca Franci}
\affiliation{School of Physics and Astronomy, Queen Mary University of London, Mile End Road, London E1 4NS, United Kingdom}
\affiliation{INAF, Osservatorio Astrofisico di Arcetri, Firenze, Italy}

\author[0000-0002-8175-9056]{David Burgess}
\affiliation{School of Physics and Astronomy, Queen Mary University of London, Mile End Road, London E1 4NS, United Kingdom}

\author[0000-0002-5608-0834]{Petr Hellinger}
\affiliation{Astronomical Institute, AS CR, Bocni II/1401, CZ-14100 Prague, Czech Republic}
\affiliation{Institute of Atmospheric Physics, The Czech Academy of Sciences, Prague, Czech Republic}



\begin{abstract}

Highly energetic, relativistic electrons are commonly present in many astrophysical systems, from solar flares to the intra-cluster medium, as indicated by observed electromagnetic radiation. However, open questions remain about the mechanisms responsible for their acceleration, and possible re-acceleration. Ubiquitous plasma turbulence is one of the possible universal mechanisms. We study the energization of transrelativistic electrons in turbulence using hybrid particle-in-cell, which provide a realistic model of Alfvénic turbulence from MHD to sub-ion scales, and test particle simulations for electrons. We find that, depending on the electron initial energy and turbulence strength, electrons may undergo a fast and efficient phase of energization due to the magnetic curvature drift during the time they are trapped in dynamic magnetic structures. In addition, electrons are accelerated stochastically which is a slower process that yields  lower maximum energies. The combined effect of these two processes determines the overall electron acceleration. With appropriate turbulence parameters, we find that superthermal electrons can be accelerated up to relativistic energies. For example, with heliospheric parameters and a relatively high turbulence level, rapid energization to MeV energies is possible.

\end{abstract}



\section{Introduction} \label{sec:intro}

Emission from many astrophysical plasma systems can be attributed, directly or indirectly, to energetic, non-thermal electrons, but the mechanisms for their acceleration are still uncertain. In solar flares up to 50\% of the released energy is carried by energetic electrons \citep[][]{Lin1971,Benz2008,Oka2018}. In the heliosphere the super-halo electron population of the solar wind between 2 -- 200 keV is a persistent feature \citep{Wang2015}. Diffuse synchrotron emission from relativistic electrons is responsible for giant radio halos and giant radio relics in the intra-cluster medium (ICM) at megaparsec scales \citep[][]{Brunetti2014,Bykov2019}. The short cooling time for energetic electrons in the ICM poses a particular problem for explaining observations, implying some mechanism for reacceleration~\citep[][]{Kang2017}.

Turbulence is a ubiquitous property of such systems and it is fundamental for the transport and energization of charged particles. The first theoretical model of particle acceleration in turbulent electromagnetic fields considered particles interacting with randomly moving scattering centers, thereby gaining energy stochastically (``second-order Fermi'' mechanism, \citet[][]{Fermi1949}). If the scattering centers have some additional coherent large-scale motion then the acceleration is more rapid (``first-order Fermi'' mechanism, \citet[][]{Fermi1954}).

In the past 70 years, since Fermi's original papers, this topic has been extensively investigated \citep[e.g.,][]{Lazarian2012}. Many calculations have been made of particle diffusion coefficients, and energization, by solving the Fokker-Planck transport equation with some characterization of the turbulent fluctuations~\citep[][]{Jokipii1966, Schlickeiser1994, Matthaeus2003}. Similarly, there have been many tests of these theories using test particle methods, i.e., advancing the equations of motion for the charged particles in prescribed turbulent fields~\citep[][]{Ambrosiano1988, Miller1997, Giacalone1999}.

Recently, the effects of current sheets and other coherent structures have been investigated using combinations of MHD and test particle simulations \citep[][]{Arzner2004,Dmitruk2004,Vlahos2019}. Self-consistent kinetic plasma simulations have led to an improved understanding of the dynamics of plasma turbulence and particle transport \citep[][]{Servidio2016,Zhdankin2019}. 

Another possible mechanism for particle acceleration in plasmas is magnetic reconnection \citep[][]{Schopper1999,Lazarian1999,Heerikhuisen2002,Kowal2012}. Although reconnection studies are often in the context of an idealized 2D geometry, recent studies have shown that magnetic reconnection and turbulence are inextricably linked \citep[][]{Servidio2012,Franci2017,Papini2019, Shay2018, Ergun2018}. \citet[][]{Drake2006}, using fully kinetic particle-in-cell (PIC) simulations, found that electrons can be efficiently accelerated in contracting magnetic islands produced by reconnecting current sheets. This is closely related to first-order Fermi acceleration, and has been used to explain energetic particles observations near the heliospheric current sheet \citep[][]{Khabarova2016}. Systems of multiple current sheets evolve to become turbulent \citep{Gingell2017}, so, in terms of particle acceleration, the relative role of magnetic structures in turbulence, compared with the closed magnetic islands found in reconnection geometries, becomes an interesting topic.

\begin{figure*}
\includegraphics[width=\textwidth]{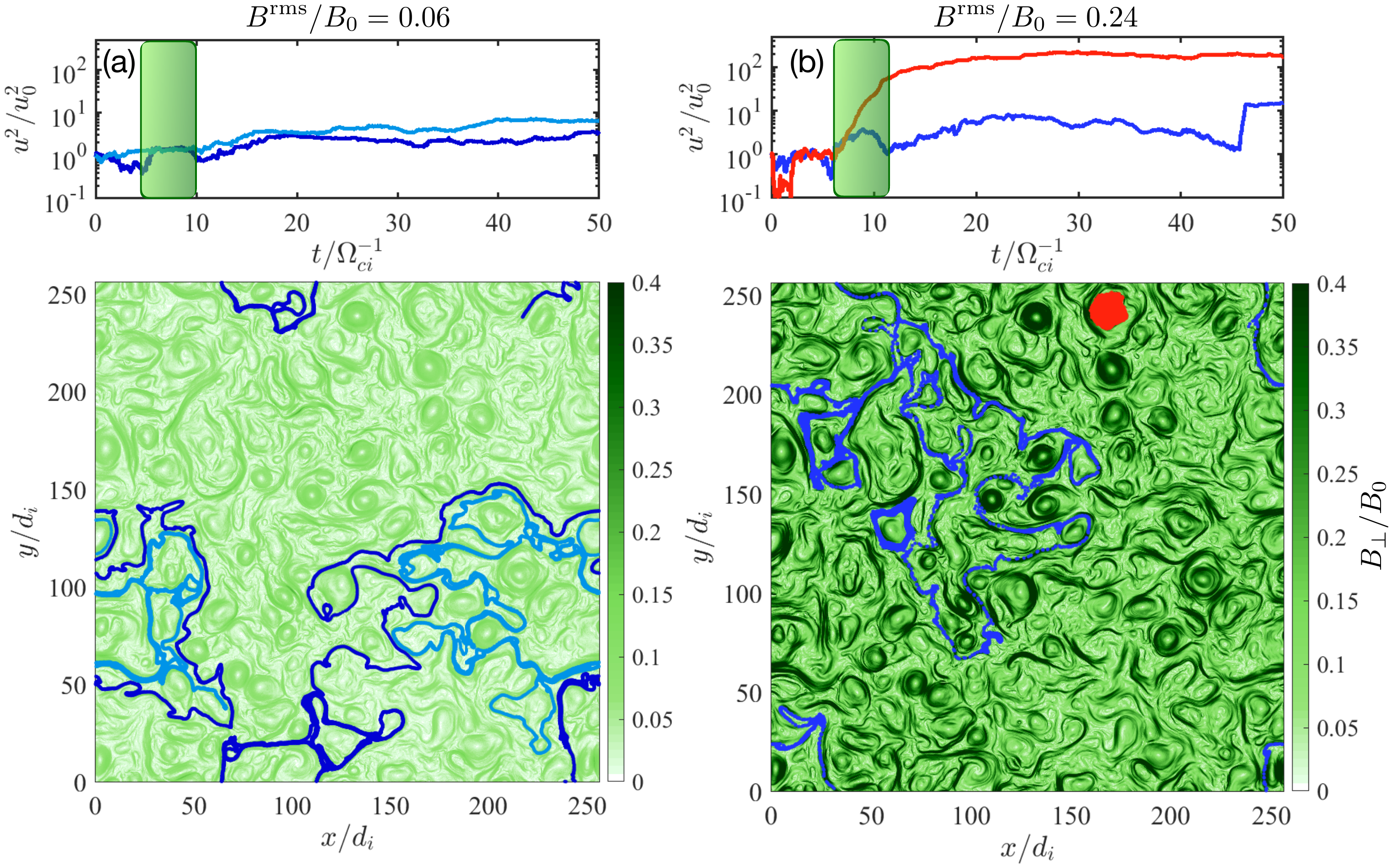}
\caption{Trajectories and energization history for electrons with initial energy corresponding to a velocity of 200 $v_A$. Plasma simulations with initial values of ${B}^{\mathrm{rms}}/B_0 = 0.06$ and $0.24$, are shown in left and right panels respectively. \emph{Top panels}: Electron energy as a function of time. \emph{Bottom panels}: Typical electron trajectories (blue, red lines), displayed for the time highlighted by the green shaded areas in the top panels. The trajectories are superimposed on the perpendicular magnetic field intensity at the mean time of the shaded area in the corresponding top panel. \label{fig:fig1}.}
\end{figure*}

Recent simulations of solar wind turbulence provide an accurate model of the plasma dynamics from MHD scales through to ion scales, where collisionless dissipation begins to be important \citep{Franci2017}. For electrons in astrophysical plasmas, the transrelativistic energy range (roughly 1 keV -- 1 MeV) corresponds to a gyroradius scale of the order of the thermal ion kinetic scales (gyroradius and inertial length). Since this energy range is important for the production of high energy electrons, it becomes interesting, for this population, to return to Fermi's original ideas and consider the relative importance of the various possible acceleration processes in turbulence.

We use hybrid PIC plasma simulations (fluid electrons and kinetic ions) and test particle modeling for the energetic electrons. Hybrid simulations provide an accurate model of the ion-scale break in the turbulence power spectrum, which corresponds roughly to the gyroradius of transrelativistic electrons. Test particle methods allow for a realistic separation of scales, compared to full PIC simulations (with kinetic electrons) which usually use a reduced ion-electron mass ratio. 

We show that the electron acceleration changes in nature for different combinations of initial energy and turbulence strength. When the turbulence amplitude is low (with respect to the background magnetic field), electrons initialised with the same energy exhibit moderate energization, and the resulting electron energy spectra are compatible with those derived analytically for a second-order Fermi mechanism only. For higher turbulence strength, a stage of fast energization is observed, and the resulting electron energy spectra cannot be modeled as only a second-order Fermi process. The fast electron energization is produced by drifts at play while particles are trapped at magnetic structures in the turbulence. This implies that the presence of large amplitude turbulence could be an important factor for explaining high energy electrons in astrophysical systems.

\section{Methods} \label{sec:methods}

The simulations consist of two stages. First, a  simulation of freely decaying, Alfv\'enic plasma turbulence is performed, which produces a fully-developed, quasi-stationary turbulent state. Thereafter, the electromagnetic fields are stored at full cadence. Next, test particle electrons are followed by advancing their equations of motion in the evolving fields obtained from the turbulence simulation. 

We use the hybrid plasma code CAMELIA \citep[][]{Franci2018b} with a 2D periodic domain of size 256 $d_i{}^2$ ($d_i = c/\omega_{pi}$ is the ion inertial length), a spatial resolution of $\Delta x =\Delta y =  0.125 \, d_i$, and a timestep for the PIC protons of $\Delta t_p = 0.01 \, \Omega_{ci}^{-1}$, where $\Omega_{ci}^{-1}$ is the inverse cyclotron frequency. Velocities are normalised to the Alfv\'en speed $v_A$.  We employ 1024 particles per cell, to keep the statistical noise to a reasonable level. Protons and electron fluid are initialised with plasma betas (ratio of kinetic to magnetic pressure) $\beta_i=\beta_e=0.5$. A magnetic guide field is imposed in the out-of-plane direction $\mathbf{B}_0 = B_0 \hat{z}$. Further details of the initialisation can be found in \citet[][]{Franci2015b}. 

\begin{figure}
\includegraphics[width=.44\textwidth]{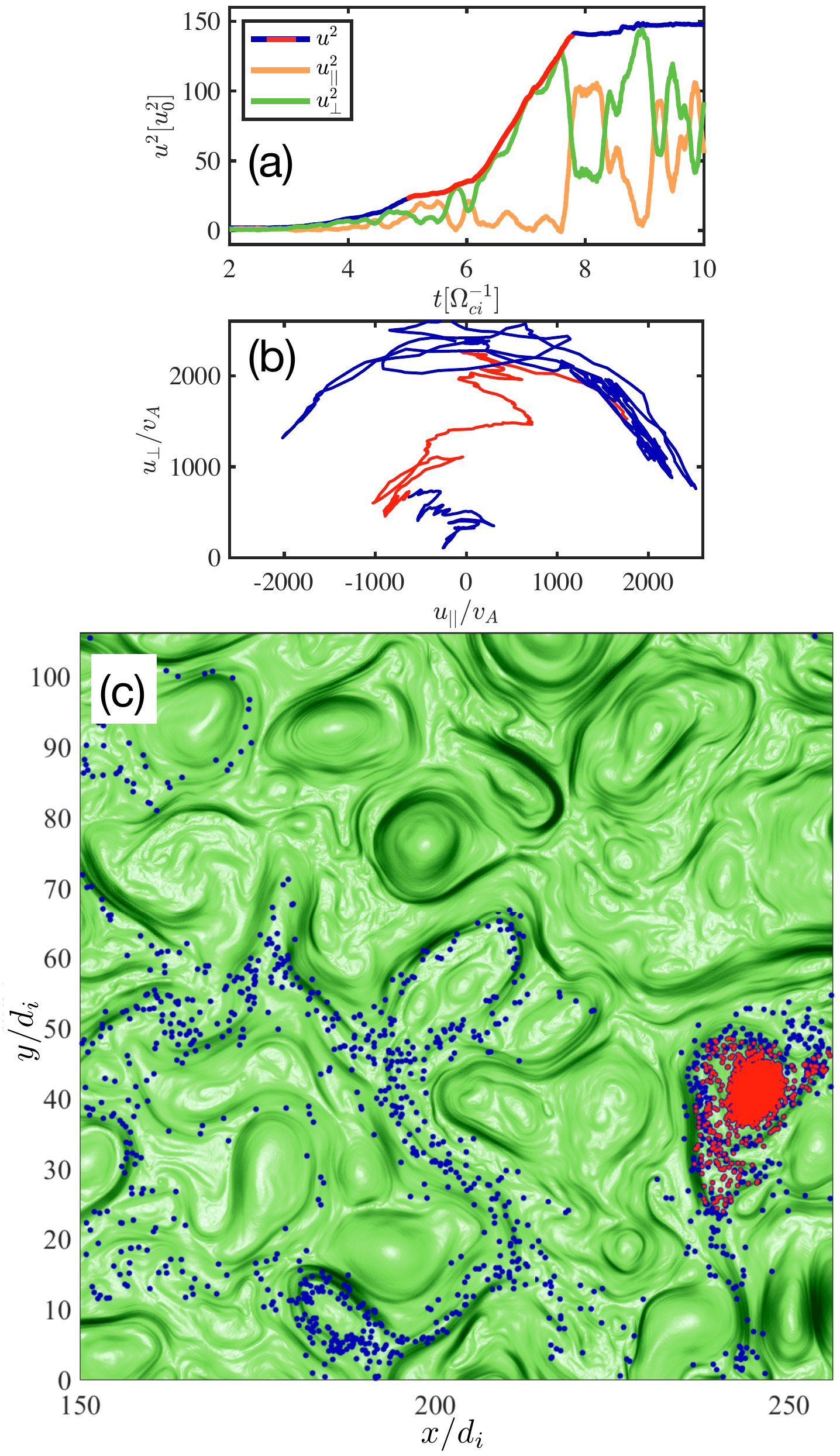}
\caption{(a): Electron energy history (blue line), with the trapped portion highlighted in red. The green and orange lines represent $u_\perp^2$ and $u_{||}^2$. (b): $u_{||}$ - $u_{\perp}$ with the portion corresponding to the trapped fragment highlighted in red. $u_{||}$ and $u_{\perp}$ have been calculated with respect to the local magnetic field. (c) Electron trajectory in the high turbulence strength case, with the perpendicular magnetic field in background. The red dots highlight the part of the trajectory in which the electron is trapped and which corresponds to fast energization.  \label{fig:fig2}} 
\end{figure}

Test-particle electrons are injected uniformly as monoenergetic velocity shells, and their motion is followed for 50~$\Omega_{ci}^{-1}$ using a relativistic Boris scheme \citep[][]{Ripperda2018} with timestep $\Delta t_e = 10^{-6} \, \Omega_{ci}^{-1}$. The ratio $c/v_A$ is 5000, typical of the solar wind. {The electromagnetic fields obtained from the hybrid PIC simulations are interpolated in space and time at the electron positions. All the electrons are initialised with energies such that their gyroradii are larger than the spatial resolution employed in the hybrid simulation.} 

The electrons are released once a turbulent cascade has fully developed, and the characteristic spectrum of the magnetic fluctuations is observed, i.e., two different power laws at scales above and below the ion-scale break, with slopes compatible with -5/3 and -3, respectively \citep{Franci2015b}.

\section{Results} \label{sec:results}

Figure~\ref{fig:fig1} shows some representative trajectories for electrons with initial energy corresponding to a velocity of 200 $v_A$ (energy 410~eV for $c/v_A =5000$). Two plasma simulations are shown with different initial values of ${B}^{\mathrm{rms}}/B_0 = 0.06$ and $0.24$ (left and right panels respectively). We will refer to these two cases as ``low'' and ``high'' turbulence strength; other cases have been studied to confirm the results presented.

The trajectories which show the least energization correspond to electrons moving across the whole simulation domain, i.e., with ``open'' trajectories, gaining and losing energy in a stochastic fashion (both trajectories in Fig.~\ref{fig:fig1}a, blue trajectory in Fig.~\ref{fig:fig1}b).

When the level of turbulent fluctuations is high, trapped (or quasi-trapped) orbits associated with a quasi-monotonic, rapid energy increase are found (Fig.~\ref{fig:fig1}b, red line). In this case, the energy gains resemble the first-order Fermi process (trapping in contracting islands) seen in simulations of magnetic reconnection \citep[e.g.,][]{Drake2006, Li2019a}. Both open and trapped trajectories coexist, and an electron may switch from one type to the other (see blue line, right panel, at $t\sim 45\Omega_{ci}^{-1}$), with implications for the mean energization discussed later.

Figure~\ref{fig:fig2} shows a zoom on another trajectory fragment. The trapped part of the orbit (red) corresponds to  a rapid energy increase (panel a), primarily due to an increase in the perpendicular velocity (panel b). Electrons may gain (or lose) energy if they have adiabatic drift motion (curvature or ``grad B'') with a component parallel to the motional electric field of the turbulent plasma motion. From panel (b), the fast energization proceeds until the particle de-traps, then the trajectory becomes open and pitch angle scattering dominates, erasing the memory of the energization process.


Trapping occurs in dynamic magnetic structures within the turbulence, which resemble magnetic vortices or flux ropes aligned in the guide field direction. Magnetic reconnection does occur, at the boundaries between magnetic vortices, but the associated electric field does not seem to give any substantial contribution to particle acceleration.

\begin{figure}
\includegraphics[width=0.47\textwidth]{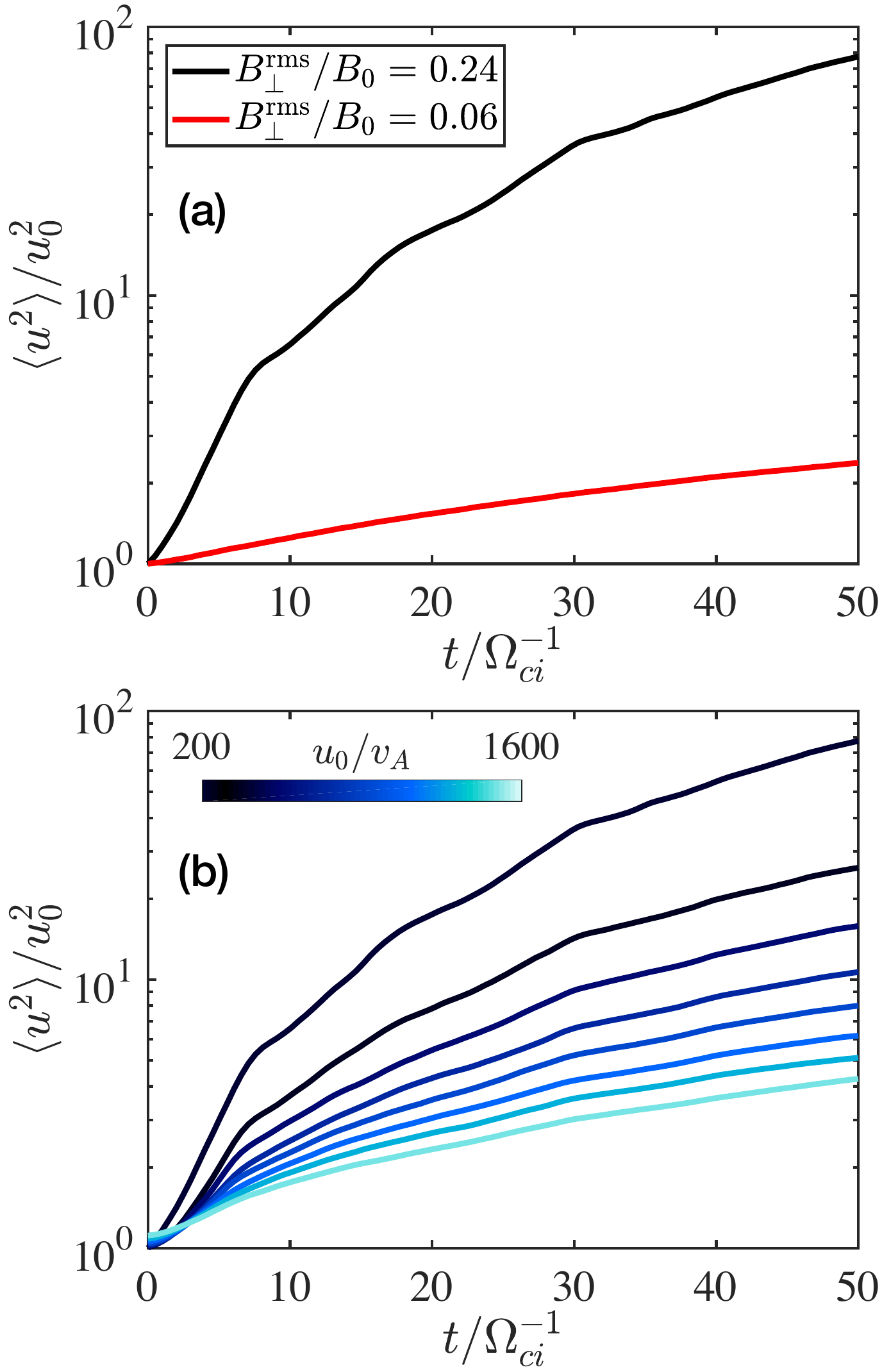}
\caption{ (a): Mean electron energy of an ensemble of electrons as a function of time in low and high turbulence strength cases (red and black lines, respectively). In both cases, the electrons are initialised with an initial energy corresponding to a speed of 200 $v_A$. (b): Mean electron energy as a function of time for the high turbulence strength case, with eight different initial electron energies corresponding to linearly spaced velocities from 200 to 1600 $v_A$ (black to light blue curves). \label{fig:fig3}}
\end{figure}

To support the indications provided by single particle trajectories, the energization of an ensemble of $10^4$ electrons is studied (Fig.~\ref{fig:fig3}a). When the turbulence strength is low (red line), the mean energy increases slowly, with only a moderate gain. In the high turbulence strength case (black line), the final energy gain is much larger, of about a factor of 100. Moreover, the electron energization proceeds in stages, i.e., until about 30 $\Omega_{ci}^{-1}$ after the electron injection, the energy gain is much faster than in the low amplitude case, but later the energy grows more slowly, similar to the low turbulence strength case.

\begin{figure}
\includegraphics[width=0.48\textwidth]{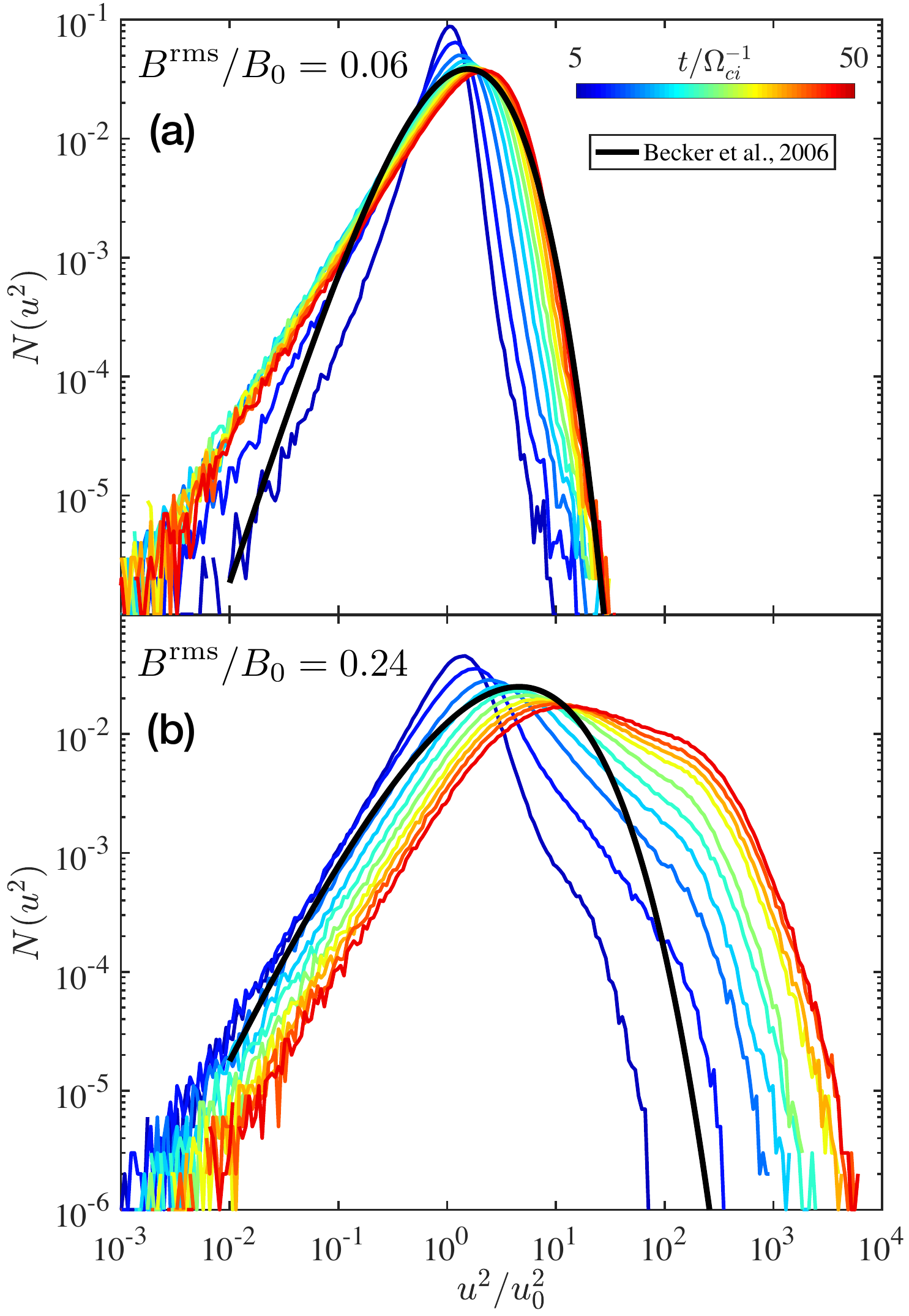}%
\caption{Electron energy spectra at various times of the simulation (colored lines). The black lines represent the analytical solution derived by \citet[][]{Becker2006}. The (a) and (b) panels correspond to the cases of low and high turbulence strength, respectively. \label{fig:fig4}}
\end{figure}

The rapid mean energy gain is due to the fast energization observed for trapped orbits (Figs.~\ref{fig:fig1} and \ref{fig:fig2}). This behavior is found to depend on the initial electron energy (Fig.~\ref{fig:fig3}b). For initial energies corresponding to a velocity of 800 $v_A$ or more, the fast regime is almost absent, and particles gain energy slowly, as in the case of low energy electrons injected in low-amplitude turbulence (Fig.~\ref{fig:fig3}a red line). This suggests that the trapping in magnetic structures is only effective for certain combinations of turbulence amplitude and initial electron energy. For example, the relative size of electron gyroradius and magnetic field gradients are likely to control how efficiently electrons are trapped. 

Figure~\ref{fig:fig4} shows the time evolution of the energy spectra (colored lines) in the low and high turbulence strength cases (panels a and b, respectively). For comparison, we also show analytical solutions for the Fokker-Plank equation considering an impulsive injection of a monoenergetic shell of particles and assuming a spectrum of fluctuations that follows a Kolmogorov scaling $P(k) \sim k^{-5/3}$ \citep[][]{Becker2006}. The aim is to illustrate the form of the solution predicted from a purely second-order Fermi process and not to attempt a quantitative fit.  It can be noted that, in the case with high turbulence strength, the electron energy spectrum is still slowly evolving (Figure~\ref{fig:fig4}b). The steady-state shape of the spectrum in the high turbulence case will be object of further investigation.  

When the level of turbulent fluctuations is low, the maximum final energy gain, as anticipated from the mean energization, is only moderate (Fig.~\ref{fig:fig4}a). The electron energy spectra qualitatively resemble the analytical prediction for second-order Fermi acceleration, with small differences possibly due to some contribution from electron trapping, or some mis-match with the assumed fluctuations for the analytical prediction. When the turbulence level is high (Fig.~\ref{fig:fig4}b), the spectra show a flattening at intermediate energies (10 to 10$^3$ $u_0^2$), and considerably higher maximum energy gains. It is clear that, in this case, the simulation spectra and the prediction of second-order Fermi do not agree in form.
{We do not find any clear sign of anisotropies in velocity space for electron Velocity Distribution Functions (VDFs), collected over the whole simulation domain, when trapping is active. These results can be related to models of first order acceleration in contracting magnetic islands, in presence of strong scattering~\citep{LeRoux2016}. Further studies will include the analysis of electron VDFs at different locations of the simulation domain (e.g., close to trapping zones, in proximity of current sheets).}

\begin{figure}
\includegraphics[width=.47\textwidth]{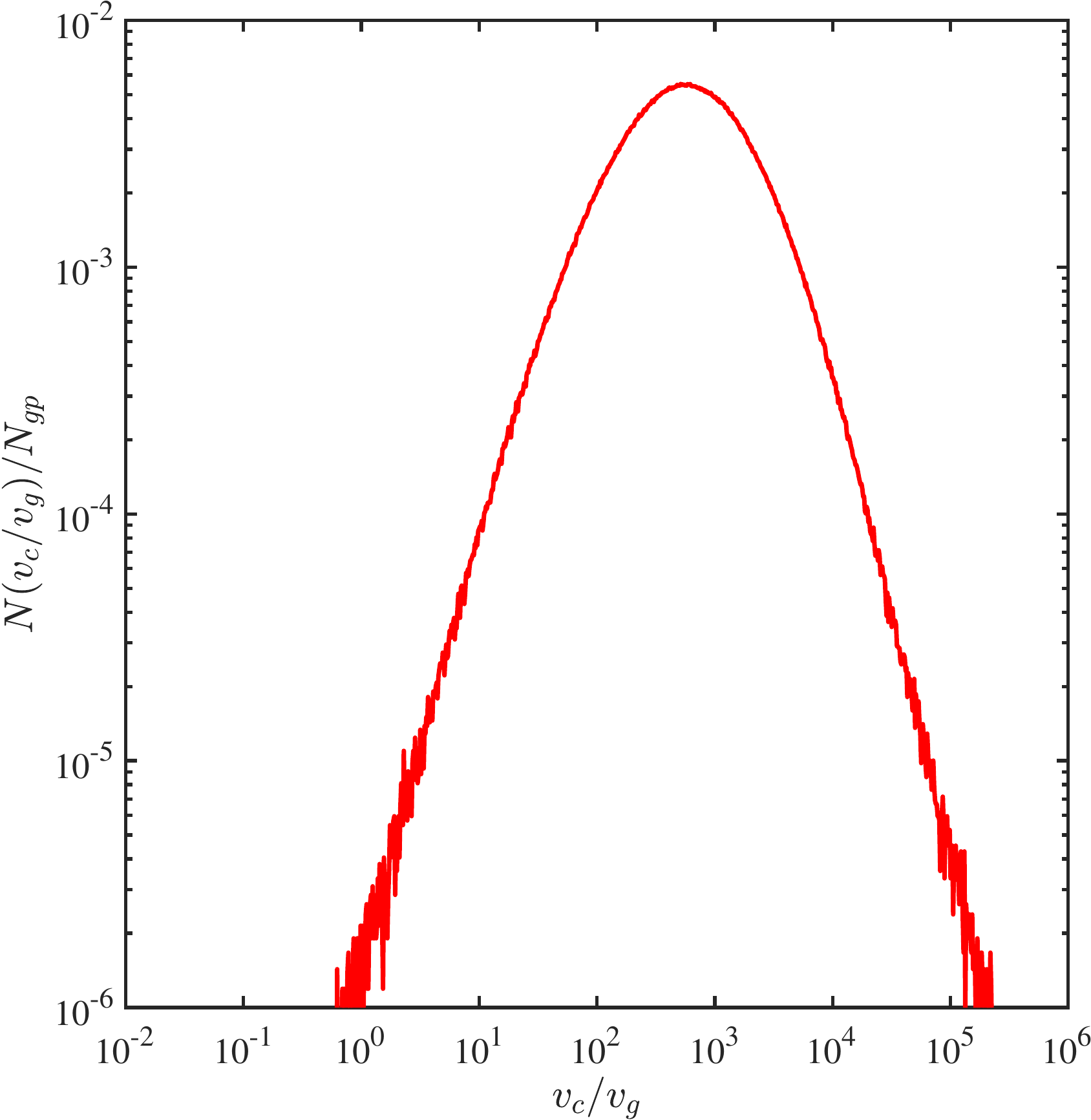}
\caption{ Distribution of the ratios between the magnetic parts of curvature and grad-B drift, computed in every point of the simulation domain. \label{fig:fig5}} 
\end{figure}

We have examined the relative magnitude of the magnetic field curvature and ``grad B'' terms for the particle drifts, and found it is consistent with curvature drift acceleration as proposed in~\citet[][]{Dahlin2016, Yang2019}. In the guiding centre limit, it is possible to define the curvature and grad-B drifts as, respectively: 
\begin{eqnarray}
\mathbf{v}_c & = & \frac{v_{||}^2}{\Omega_{ce}} \mathbf{b} \times \kappa \label{eq:eq1} \\
\mathbf{v}_g & = &\frac{v_{\perp}^2}{2 \Omega_{ce}} \mathbf{b} \times \frac{\nabla B}{B} . \label{eq:eq2}
\end{eqnarray}
Here, $v_{||}$ and $v_{\perp}$ are the parallel and perpendicular electron speed with respect to the local magnetic field and $\Omega_{ce} = eB/\gamma m_e c$ is the electron cyclotron frequency. $B$ indicates the magnetic field  magnitude and $\mathbf{b} = \mathbf{B}/B$. $\kappa = \mathbf{b} \cdot \nabla \mathbf{b} $ is the magnetic field curvature vector. 

To support the evidence that curvature drift acceleration is dominant with respect to grad-B drift, we computed the ratio between the magnetic part of each drift (i.e., assuming that $v_\perp/v_{||} \approx 1$), in each point of the simulation domain. Figure~\ref{fig:fig5} shows the resulting distribution of these ratios throughout the simulation domain, normalised to the total number of gridpoints in the simulation, $N_{gp}$. As it can be seen, the curvature drift term systematically exceeds the grad-B one.


\section{Conclusions} \label{sec:conclusions}

Using a combination of hybrid PIC and test particle simulations we have shown that, for high turbulence strength, transrelativistic electrons can be rapidly and efficiently accelerated by turbulence. For example, for turbulence with ${B}^{\mathrm{rms}}/B_0 = 0.24$, with electron energy injection at about 400~eV, the mean energy reaches about 40~keV (Fig.~\ref{fig:fig3}a, black line), and the maximum energy attains over 1~MeV (Fig.~\ref{fig:fig4}b).

Turbulence strength is found to control the mechanisms of electron acceleration. For low turbulence amplitude, electrons are found to be only moderately energized, consistent with the standard theory of stochastic acceleration \citep[][]{Fermi1949}. On the other hand, when the turbulence amplitude is high, electrons are energized more efficiently, with a fast, ``injection'' stage due to trapping and subsequent acceleration in turbulent structures. There is evidence that the energization is dominated by curvature drift acceleration \citep[][]{Dahlin2016,Yang2019}. At higher initial energies, the fast energization stage is not dominant. Hence, the efficiency of trapping depends on both the electron energy and the level of turbulent fluctuations. 

The final electron energy spectra are the result of the interplay between fast energization in trapped trajectories  and stochastic acceleration in open trajectories, as it can be seen in Figure~\ref{fig:fig4}, where a qualitative comparison between the energy spectra resulting from our simulations and those calculated in \citet{Becker2006} was performed.  In any realistic system, loss processes would have to be accounted for before predicting actual energy spectra.

We note that the mixture of trapped and open trajectories of accelerated electrons implies that the behavior of the system cannot be properly modeled by a single diffusion coefficient, since the diffusion regime (i.e., anomalous or normal) appears heterogeneous.

The employed simulation method gives a realistic separation of scales for energetic electrons in a proton-electron plasma, with the turbulence properties accurately modeled down to ion scales, as it is extensively shown in previous literature. An important limitation of this approach is that the feedback of test-particle electrons on the turbulent electromagnetic fields is neglected. Combining hybrid PIC and test-particle simulations is complementary to the fully kinetic PIC approach, that has its own computational limitations, such as the use of reduced ion to electron mass ratios~\citep[see, for example,][]{Li2019b}. Our approach is also complementary to fully kinetic PIC simulations of turbulent acceleration in an electron-positron plasma \citep{Comisso2018,Comisso2019}, whose results are similar to those reported here, e.g., the evolution of the energy spectrum. However, in the case of \citet{Comisso2018} the energy gain from reconnection electric fields is important, in contrast to energization via trapping in magnetic structures, as found here.

The results presented here are broadly relevant for the acceleration (and reacceleration) of energetic electrons in the heliosphere and in other astrophysical systems. For example, recently, enhancements of flux of energetic particles have been observed in conjunction of magnetic flux tubes the inner heliosphere~\citep{Bandy2019}. Examples of other astrophysical systems where these results may be relevant are the giant radio halos of galaxy clusters \citep[][]{Bykov2019} and relativistic jets of active galaxies~\citep[][]{Alves2018}. The dependence of electron energization on the level of turbulent fluctuations may be important for systems that naturally produce an increase of turbulence strength, such as downstream of shocks \citep[][]{Kang2017}.

The work presented here employed two-dimensional simulations. The efficiency of electron trapping is likely to be affected by a more realistic, 3D geometry. Future studies will consider the effect of the reduced 2D geometry of our simulations, although it has been shown that similar simulations give comparable results (in terms of spectral and intermittency properties) to 3D simulations and spacecraft observations with the same plasma conditions \citep{Franci2018, Franci2019Ar}. Finally, the domain used limits the size of the largest-scale magnetic islands, and restricts the development of the large-scale inertial range.

\begin{acknowledgments}

This work was performed using the DiRAC Data Intensive service at Leicester, operated by the University of Leicester IT Services, which forms part of the STFC DiRAC HPC Facility (www.dirac.ac.uk), under the project “dp031 Turbulence, Shocks and Dissipation in Space Plasmas”. The equipment was funded by BEIS capital funding via STFC capital grants ST/K000373/1 and ST/R002363/1 and STFC DiRAC Operations grant ST/R001014/1. DiRAC is part of the National e-Infrastructure.
This research utilised Queen Mary's Apocrita HPC facility, supported by QMUL Research-IT. http://doi.org/10.5281/zenodo.438045. This research was supported in part by the National Science Foundation under Grant No.\ NSF PHY-1748958.
\end{acknowledgments}


\end{document}